\documentclass[conference,letterpaper,10pt]{IEEEtran}
\usepackage[T1]{fontenc}
\usepackage[latin1]{inputenc}
\usepackage{float}
\usepackage{amsmath}
\DeclareMathOperator*{\argmax}{arg\,max}
\usepackage[noadjust]{cite}
\usepackage{amssymb}
\usepackage{enumerate}
\usepackage{amsthm}
\usepackage{graphics,epsfig,psfrag}
\usepackage{subfigure,epsf,pstricks,subfloat}
\usepackage{bm}
\usepackage{url}
\usepackage{gensymb}

\usepackage{comment}
\floatstyle{ruled}
\newfloat{algorithm}{tbp}{loa}
\floatname{algorithm}{Routine}

\usepackage{algorithmic}
\algsetup{linenodelimiter= }

\title{Likelihood of Cyber Data Injection Attacks to Power Systems} 
\author{
\IEEEauthorblockN{Yingshuai Hao, \textit{Student Member, IEEE,}  Meng Wang, \textit{Member, IEEE}, Joe Chow, \textit{Fellow, IEEE} 
\IEEEauthorblockA{Department of Electrical, Computer, and Systems Engineering\\
Rensselaer Polytechnic Institute, Troy, NY, 12180, USA\\
Email:   \{haoy2,wangm7,chowj\}@rpi.edu}
 }
}

\usepackage{array}
\usepackage{multirow}
\usepackage{graphicx}
\usepackage{epstopdf}

\begin{document}

\maketitle \thispagestyle{empty} \pagestyle{empty}

\begin{abstract}
Cyber data attacks are the worst-case interacting bad data to power system state estimation and cannot be detected by existing bad data detectors.
% Recent research on cyber data attacks in smart grids is centered on the  protection of key measurement to avoid these attacks, as well as the development of new attack detection methods.  What is missing is the analysis of the likelihood of cyber data attacks in a power grid. 
%Such analysis is essential to study the vulnerability to cyber attacks. 
In this paper, we for the first time analyze %the frequency and 
the likelihood of cyber data attacks by characterizing  the actions of a malicious intruder. 
We propose to use Markov decision process to model an intruder's strategy, where the objective is to maximize the cumulative reward across time. % where the 
%immediate reward results from the price gap between different load states.
%
Linear programming method is employed to find the  optimal attack policy from the intruder's perspective.   Numerical experiments are conducted to  study the intruder's attack strategy in test power systems. 
\end{abstract}
\begin{IEEEkeywords}
cyber data attacks, Markov Decision Process, state estimation, power systems.
\end{IEEEkeywords}

\section{Introduction}\label{sec:intro} 
 
State  estimation \cite{AE04} solves for power system states from obtained measurements. The correct estimation of systems states is vital for the reliable operation of  %the monitoring and control of 
power systems.  
Since  bad data 
%(erroneous measurements) 
can result in significant errors in  the outcome of state estimation and potentially lead to catastrophic consequences,   %due to  malfunction of the devices 
the detection and identification of bad data has been extensively studied \cite{CA06,HSKF75,MG83,MS71,XWCT13} in state estimation. 
  
 % Despite the benefits to intelligent power system operation, 
 The integration of cyber infrastructures in future smart grids inevitably increases the possibility of cyber attacks. 
Cyber data attacks, firstly studied in \cite{LNR11}, 
%are viewed as ``the worst interacting bad data injected by an adversary''\cite{KJTT10}. 
means that a malicious intruder with system configuration information simultaneously manipulates   multiple  measurements and the injected errors cannot be detected by any bad data detector. % and would introduce significant errors to the output of state estimators. 
%These worse-case cyber attacks are firstly studied in \cite{LNR11}.  Because   the removal of affected measurements would make the system %unobservable, %these worst-case cyber attacks are firstly studied in \cite{LNR11} and 
%these attacks are termed as   ``unobservable attacks''\footnote{The term ``unobservable'' is used in this sense throughout the paper.} in %\cite{KJTT10}.  
%
%It is shown in \cite{LNR11} that attackers with the system topology information can inject alter the measurements of a few meters and lead to significant error to

%It was first discovered in \cite{LNR11} that some , these attacks are termed as ``unobservable attacks'' in \cite{KJTT10}
State estimation in the presence of cyber data  attacks has attracted much research attention recently  \cite{BRWKNO10,DS10,KJTT10,LEDEH14,LNR11,SJ13,STJ10,TKPC11}. Some work focused on the identification 
% and protection  
of a small number of key measurement units 
%to prevent any  without hacking the protected units. .
such that if those units are protected from an intruder, the intruder cannot launch a successful cyber data attack \cite{BRWKNO10,DS10,KP11}.
% \cite{STJ10,DS10} proposed security indices for state estimators based on the requirements to launch a cyber data attack.  
A few recent work \cite{SJ13,LEDEH14,WGGCFSR14} considered the detection of cyber data attacks in a power system with Supervisory Control and Data Acquisition (SCADA) systems or phasor measurement units (PMUs) and various detection methods have been proposed. % as well as detectors designed for attacks in the observable regime \cite{KJTT10}. The research on the detection of unobservable attacks is still limited. Refs. \cite{SJ13,LEDEH14} proposed different methods %a detection method based on statistical learning 
%to detect unobservable attacks in Supervisory Control and Data Acquisition (SCADA) system. % of the power system.
%The method in \cite{SJ13} %provides no theoretical guarantee on their method which 
%relies critically on the assumption that the measurements at different time instants are i.i.d. samples of random variables. This assumption might not hold when the system is experiencing some disturbance.  %\cite{LEDEH14} proposed a  method that exploits the temporal correlations in the actual measurements to detect unobservable attacks in SCADA system. 
%Refs.~\cite{SJ13,LEDEH14} studied  the detection of data attacks in  %and  proposed detection methods when an intruder attacks a different set of measurements at each time instant. %, and no theoretical analysis of the detection performance is provided in  \cite{LEDEH14}.
%and Ref.~\cite{WGGCFSR14} considered the detection of data attacks in  (PMUs).
%and focused on the scenario that an   intruder attacks a few PMUs and  injects  data attacks to the same set of PMUs constantly. 
%Ref.~\cite{WGGCFSR14} proposed a new detection method and provided its theoretical detection guarantee. 

What is missing in the literature of cyber data attacks is the analysis of the frequency of data attacks in smart grids and the likelihood of attacks at a given system state. 
%Such analysis is essential to study the vulnerability of power systems to cyber data attacks
%, as well as to predict the frequency of future attacks. 
% and evaluate the benefits of incorporating various attack detection methods in state estimation. 
This paper takes the first step to  analyze the likelihood of data attacks from the intruder's perspective. We consider a scenario that  if a cyber data attack is detected by the operator,   the affected measurement units will be protected for some time. Thus, the intruder's current action  affects its future available actions.
%, while power systems evolve  across time, and future system states are unknown to the intruder. %Then  the major challenge for an intruder to develop an attack strategy results from the fact that  power systems evolve  across time, and future system states are unknown to the attacker, while the intruder's current action may affect its available actions in the future. To address this issue, 
To address such challenge in  developing an attack strategy, we propose to use Markov Decision Process (MDP) \cite{MP94} to  model the intruder's attack decision across time. %We apply linear programming \cite{FR03} to solve the resulting MDP formulation. The solution
The solution to the resulting MDP
is a mapping from system states to the intruder's actions (attack or not, which bus to attack and how much error to inject). Numerical experiments are carried on PJM 5-bus system to verify the proposed approach and study the likelihood of cyber attacks in such systems. 
%Although the paper is centered on data attacks to PMU measurements, the proposed method can be easily extended to study data attacks in SCADA systems. 
%To the best of our knowledge, this paper provides the first study on the frequency of cyber data attacks.

The rest of the paper is organized as follows. We motivate the problem and introduce cyber data attacks and MDP in Section \ref{sec:model}. We formulate the intruder's attack strategy as an MDP and introduce its solution method in Section \ref{sec:method}.  Section \ref{sec:simu} records our numerical study on an example network. We conclude the paper in Section \ref{sec:con}. 

\section{Problem Motivation and Background} \label{sec:model}

We first introduce the definition and existing work on cyber data attacks in Section~\ref{sec:cyber}. We then motivate and introduce the problem of likelihood analysis of cyber data attacks in Section~\ref{sec:problem}. One main contribution of our paper is to model this problem as a Markov Decision Process (MDP).  MDPs are introduced in Section \ref{sec:MDP}.
%Firstly we discuss the cyber data attack in power system and present our main contribution. Then we describe the event sequence of data attack and review the framework of MDP employed to model our problem. 
\subsection{Cyber Data Attacks in Power Systems}\label{sec:cyber}
In a power system, the state is usually represented by bus voltage magnitudes $\bm{V}\in \mathcal{R}^n$ and angles $\bm{\theta}\in [-\pi,\pi]^n$, where $n$ is the number of buses. 
State estimation \cite{AE04} solves for system states from the obtained measurements.
% such as power injections, line power flow and voltage magnitudes. 
Under the AC measurement model, the measurements $\bm{z}$ can be expressed as a nonlinear function of state variables $\bm{x}=(\bm{V}$,$\bm{\theta}$):
\begin{equation}
\bm{z}=h(\bm{x})+\bm{\omega},
\end{equation}
where $\bm{\omega}$ represents the random measurement noise.

%One example is that the real power flow from bus $a$ to bus $b$ via line $ab$ is
%\begin{equation}\label{eqn: real power}
%P_{ab}=\frac{V_a^2}{Z_{ab}}\cos \theta_{Z_{ab}}-\frac{V_a V_b}{Z_{ab}}\cos(\theta_a-\theta_b+\theta_{Z_{ab}}),
%\end{equation}
%where $V_a,\theta_a, V_b, \theta_b$ corresponds to the voltage magnitude and angle of bus $a$ and $b$ respectively, $Z_{ab}$ is the impedance of line $ab$, $Z_{ab}=R_{ab}+jX_{ab}$, where $R_{ab}$ and $X_{ab}$ represent the resistance and reactance of line $ab$ respectively. $\theta_{Z_{ab}}$ is the impedance angle, $\theta_{Z_{ab}}=\arctan(X_{ab}/R_{ab})$. 

%When the resistance of line $ij$ can be ignored, formula (\ref{eqn: real power}) can be further expressed as 
%\begin{equation}\label{eqn: reduced real power}
%P_{ij}=\frac{V_i V_j}{Z_{ij}}\sin(\theta_i-\theta_j).
%\end{equation}
In the AC state estimation, the state variables are determined from the weighted least square optimization problem:
\begin{equation}
\bm{\hat{x}}=\arg\min(\bm{z}-h(\bm{x}))^T\cdot\bm{R}^{-1}\cdot(\bm{z}-h(\bm{x})),
\end{equation}
where $\bm{R}$ is the covariance matrix of measurement noise $\bm{\omega}$.

Malicious intruders can hack the measuring devices and inject interacting errors to the measurements. If they have sufficient system information and choose the errors $\bm{e_z}$ satisfying
\begin{equation}\label{eqn: measurements with errors}
\begin{split}
\bm{z}+\bm{e_z} & =h(\bm{x'})+\bm{\omega}\\
& =h(\bm{V}+\bm{e_V},\bm{\theta}+\bm{e_\theta})+\bm{\omega},
\end{split}
\end{equation}
where $\bm{e_V}$ and $\bm{e_\theta}$ represent the resulting errors on state variables $\bm{V}$ and $\bm{\theta}$ respectively, the manipulated measurements cannot be detected by existing bad data detectors. In this case, instead of correctly estimating state variables ($\bm{V},\bm{\theta}$), the operator would obtain a wrong estimate $(\bm{V}+\bm{e_V},\bm{\theta}+\bm{e_\theta})$.

%Bad data processing is carried out during state estimation to detect and exclude bad measurements.  A malicious intruder with system information can inject interacting errors to multiple measurements simultaneously \cite{LNR11}. With appropriate manipulation of measurements, such cyber data attacks cannot be detected by existing bad data detectors and can result in significant errors in the output of a state estimator.

%Phasor measurement units (PMUs) \cite{PT08} directly measure bus voltage phasors and line current phasors at synchronized sampling instants. At time $t$, the phasor measurements are linearly related with corresponding state variables as follows.
%Consider a power grid with $m$ buses and PMUs deployed on some buses to measure the voltage and current phasors, here we define the  bus voltage phasors as the state variables and with these measurements, the model of power system can be given by:
%
%where $e_{I_u}$ and $e_{V_u}$ denote the vectors of inject errors into the measured current and voltage phasors respectively; $e_V$ is the injected error on the vector of $m$ bus voltage phasors. In this case, the contaminated data can bypass the traditional bad data detection techniques and theoretically arbitrary error can be injected in the measurements and estimated system state. 

Since such cyber data attacks cannot be identified by   bad data detectors, 
%Existing approaches include %identifying and 
many efforts have been devoted to identify and protect  a small number of key measurement units such that an intruder cannot inject unobservable attacks without hacking protected %measurement 
units  \cite{BRWKNO10,DS10,KP11}. % as well as detectors designed for attacks in the observable regime \cite{KJTT10}. The research on the detection of unobservable attacks is still limited. Refs.
A few recent work \cite{SJ13,LEDEH14,WGGCFSR14} considered the detection of data attacks and various detection methods have been proposed.
% for cases where an intruder attacks a different set of measurements at each time instant \cite{SJ13,LEDEH14}, as well as cases in which an intruder attacks the same set of devices constantly \cite{WGGCFSR14}. 

The potential financial risks of cyber data attacks are studied in \cite{XMS11} and \cite{JKT14}, where the congestion pattern is defined as the set of congested lines.  By injecting false data without being detected, the intruders could change the congestion pattern and thus change the locational marginal price (LMP). The intruders can obtain financial reward from the resulting change in LMP. 

In this paper, we restrict our attention to attacks that satisfy (\ref{eqn: measurements with errors}) and result in a change of the system congestion pattern. By launching a data attack, if a line's real power is wrongly estimated to exceed its capacity while it actually is not, or the power is wrongly estimated to below its limit while the line is actually congested, then the intruder can gain a reward from the attack.

\subsection{Likelihood of Cyber Data Attacks}\label{sec:problem}
One important question that has not been considered before is the analysis of the likelihood of cyber data attacks at a given operating state of power systems. %This paper is centered on the analysis of the attack likelihood.  
 %This paper is focused on the 
%the frequency of data attacks, as well as the likelyhoold of cyber data attacks at a given state. import to identfiy the vulnarbity , protect some lines, operate away from bad states. 
%This is the focus of this paper. 
% Attacks can happen during either  PMU sampling or  data transmission to control center.  
%within a certain period of time. 
%One example of attack reward is the financial profit in electricity market. As shown in \cite{JTT12,JKL14}, if the data attacks successfully mislead the operator to identify an uncongested transmission line as congested,
%the electricity price can increase significantly. 

In this paper, we act as an intruder and aim to find the optimal attack strategy from an intruder's perspective.
We assume that an intruder can obtain a reward from a change of the congestion pattern by injecting false data without being detected.  The intruder aims to maximize the cumulative reward. If a cyber data attack, however, is detected by detection methods such as \cite{SJ13,LEDEH14,WGGCFSR14}, we assume the intruded measuring devices will be protected for some time so that an intruder cannot change the measurements of these devices during the period. %This is a reasonable assumption.  
A detected attack, therefore, can limit the intruder's future actions and thus reduce the future reward. Since the state of a power system changes across time, and the future state is unknown to the intruder, it needs to decide  when and which buses to attack to maximize the total rewards based on its current estimate.  We employ Markov Chains \cite{MP94} to model the evolution of power system states and formulate the intruder's decision process as a Markov Decision Process \cite{MP94}. The solution of resulting MDP is a mapping from system states to the intruder's actions.

\subsection{Markov Decision Processes (MDPs)}\label{sec:MDP}
An MDP is a mathematical framework employed to model the  decision-making process in stochastic environments. In this framework, the system is modeled via a series of states $S$. Each state $s\in S$ has an associated set of actions $A(s)$. In time step $t$, a decision is made based on the system's current state $s_t\in S$, and an action $a_t\in A(s_t)$ is chosen to conduct. The cost of taking action $a_t$ at state $s_t$ is $G(s_t,a_t)$. Then following the state transition probability distribution, the system transits to a new state $s_{t+1}$ with a probability of $P(s_{t+1}|s_t,a_t)$. A reward $R(s_{t+1}|s_t, a_t)$ is received from the state transition. 
As the system evolves, a sequence of rewards is obtained. 
The aim for decision-makers is to choose sequential actions that yield maximal expected rewards over the total decision-making horizon. The MDP problem can be solved by numerous methods, like value iteration, policy iteration and linear programming approaches discussed in \cite{MP94}. 

%The cumulative reward from time step $t=0$ to $T$ is $\sum_{t=0}^{T}\gamma^t(R(s_{t+1}|s_t, a_t)-G(s_t,a_t))$, where $r$ is the discount factor for future reward.
%Mathematically, an infinite-horizon MDP is defined as a 5-tuple ($S,A,P,R,\gamma$):
%$S= \{s_1, s_2,\dots, s_n\}$ represents the set of system states; $A$ is the set of actions, each state $s$ has an associated set of actions $A(s)$. 
%$P(s'|s,a)$ is the transition probability that the system transits to state $s'$  after taking action $a$ in state $s$.  
%$R(s'|s,a)$ represents the received reward as a result of state transition from state $s$ to $s'$ with action $a$. $\gamma \in(0,1)$ is the discount factor for future rewards. The goal is to find the optimal actions that maximize the cumulative expected net rewards.
%%, which is 
%% \begin{equation}\label{eqn:general}
%%E\left[\sum_{t=0}^{T}\gamma^t (\sum_{s_{t+1}}R(s_{t+1}|s_t,a_t)\times P(s_{t+1}|s_t,a_t)-G(s_t,a_t))\right].
%% \end{equation} 

\section{Problem Formulation and Solution}\label{sec:method}
In Section \ref{sec: problem formulation}, the problem is formulated from the perspective of attackers and the strategy of cyber data attacks is modeled as an infinite-horizon MDP. The solution method of resulting MDP is discussed in Section \ref{sec: MDP solution}.
\subsection{Problem Formulation}\label{sec: problem formulation}
%We first describe   the proposed MDP model of cyber data attacks in power systems.  
\subsubsection{States and Time Steps}	
%The actual power flow of target lines determines how much error to be injected in order to manipulate their estimated state to congestion.  Considering in normal operation the magnitude of bus voltages and power factor are constrained within certain scopes, we define the lines' current as the variable to determine the injected error. Moreover, the states of PMUs determine whether the measurement data is accessible to intruders, therefore the system state can be represented by states of PMUs and lines' current. 

%Since attack actions and rewards depend on line real powers and the states of measuring devices, as will be shown later, and the real power is a function of bus voltage magnitudes and angles,
 
Here we employ bus voltage magnitudes, angles and the states of measuring devices together as system states in an MDP. Because the measurements contain noise and an intruder may have limited knowledge of the states of a power system, we use discrete states to model the intruder's estimate of actual power system states. 
%Due to the uncertainty from measurement noise, we define the state of each system variable corresponds to not one certain point but an interval containing its measurement. 
For example, let $V_i$ denote the voltage magnitude of bus $i$, $V_i^{\max}$ and $V_i^{\min}$ denote the upper bound and lower bound of $V_i$ respectively. $\Delta V_i=V_i^{\max}-V_i^{\min}$. $n_V$ denotes the number of discrete states in the range. We define the state of the voltage magnitude of bus $i$ as
%Given the continuity of lines' current, firstly 
%we discretize the ratio between the actual current and the allowable maximal current into $n$ intervals, then the state of $i$th transmission lines is denoted as:
\begin{equation}\label{equ: voltage magnitude state}
\bar{V}_i=q/(n_V-1),\ q\in \{0,1,\cdots,n_V-1\},
\end{equation}
if $V_i \in \left[V_i^{\min}+q\times \tfrac{\Delta V_i}{n_V-1},V_i^{\min}+(q+1)\times \tfrac{\Delta V_i}{n_V-1}\right)$.
Similarly, let $\theta_i$ denote the voltage angle of bus $i$, $\theta_i^{\max}$ and $\theta_i^{\min}$ denote its upper bound and lower bound respectively, $n_{\theta}$ denote the number of discrete states. $\Delta \theta_i=\theta_i^{\max}-\theta_i^{\min}$. We say the state of the voltage angle of bus $i$ is
\begin{equation}
\bar{\theta}_i=q/(n_\theta-1),\ q\in \{0,1,\cdots,n_\theta-1\},
\end{equation}
if $\theta_i \in \left[\theta_i^{\min}+q\times \tfrac{\Delta \theta_i}{n_\theta-1},\theta_i^{\min}+(q+1)\times \tfrac{\Delta \theta_i}{n_\theta-1}\right)$.
%The state of voltage angle of bus $i$ $\theta_i$ shares a similar definition. We define there are $n_\theta$ distinct intervals, and if $\theta_i$ belongs to $\left[\theta_i^{\min}+q\times \tfrac{\Delta \theta_i}{n_\theta-1},\theta_i^{\min}+(q+1)\times \tfrac{\Delta \theta_i}{n_\theta-1}\right)$, where $\Delta \theta_i=\theta_i^{\max}-\theta_i^{\min}$, then its state $\bar{\theta}_i$ is
%\begin{equation}\label{equ: voltage angle state}
%\bar{\theta}_i=q/(n_\theta-1),\ q\in \{0,1,\cdots,n_\theta-1\}.
%\end{equation}
The state of the $j$th measuring device is denoted by a variable $\bar{U}_j$:
\begin{equation}
\bar{U}_j=
\begin{cases}
1 & \text{$j$th device is open to attack},\\
0 & \text{$j$th device is protected from intrusion }.		
\end{cases}
\end{equation}
If a measuring device is protected, then an intruder cannot change any measurements of that device. Otherwise, an intruder can change partial or all measurements of that device. 

%The sequence of events is shown in Fig.~\ref{fig:sequence}. 
\begin{figure}	
	\centering
	\includegraphics[ width=0.5\textwidth]{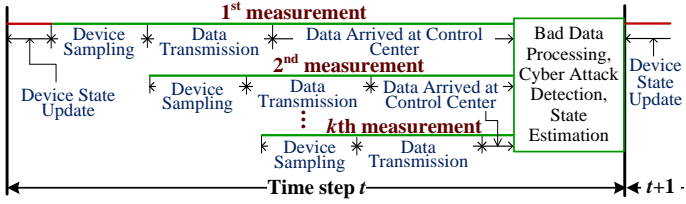}
	\caption{Event sequence with cyber data attacks. An intruder changes the observations of measuring devices to mislead the operator.}\vspace{-5mm}
	\label{fig:sequence}
\end{figure}

We consider a discrete-time system, and the time step is set as the duration between two consecutive instants of state estimation. The sampling rate of measuring devices can be higher than the state estimation frequency, as shown in Fig.~\ref{fig:sequence}. Attacks can happen during either  device sampling or  data transmission to control center. We assume if intruders decide to attack a device during step $t$, they need to change all the measurements of that device in the time step. Otherwise, the attack can be easily detected by comparing consecutive measurements. 
%PMUs measure the phasors at the sampling rate of usually 30 or 60 samples per second. 
%The control center usually estimates the system state per 1 to 5 seconds, and during the period PMUs complete tens of samplings, as shown in Fig.~\ref{fig:sequence}. The time step here is set as the state estimation period. The states of the discrete-time system are characterized after PMU sampling.   
If a power system has $M$ buses and $N$  devices,  the state $s_t$ at time step $t$ is % denoted as:
\begin{equation}\label{eqn: system state}
	\begin{split}
	s_t&= \left[\bar{\bm{V}}(t),\bar{\bm{\theta}}(t),\bar{\bm{U}}(t)\right]\\
	&=\left[\bar{V}_1(t),\cdots,\bar{V}_M(t),\bar{\theta}_1(t),\cdots,\bar{\theta}_M(t),\bar{U}_1(t), \cdots,\bar{U}_N(t)\right],
	\end{split}
\end{equation}
where $\bar{V}_i(t)$, $\bar{\theta}_i(t)$ and $\bar{U}_j(t)$ represent the states of voltage magnitude and angle of bus $i$ and the state of $j$th device at time step $t$ respectively. 

%Given that in practice the measurement data of PMUs are reported to the control center at a specified number of samples per second and the cyber data attack is to manipulate the samplings of PMUs by following (\ref{eqn:attack}), here we set the time step interval equal to the sampling period of PMUs.
\subsubsection{Actions, Rewards and Costs}
Since the reward results from a change in the congestion pattern, in order to change the line power, the intruder needs to inject errors to the estimate of voltage phasors of incident buses. Let $e_{V_i}$ and $e_{\theta_i}$ denote the injected errors to the voltage magnitude and angle of bus $i$ respectively. To make the problem tractable, we assume $e_{V_i}$ and $e_{\theta_i}$ can only be multiples of $\frac{\Delta V_i}{n_V-1}$ and $\frac{\Delta \theta_i}{n_\theta-1}$  respectively, and the resulting estimates of system variables still lie in individual allowable range. In this case, there are totally $n_V\times n_\theta$ available ways to inject errors to one bus voltage phasor. Note that in order to pass the bad data detection, given $\bm{e_V}$ and $\bm{e_\theta}$, an intruder needs to choose the injected errors $\bm{e_z}$ to measurements according to (\ref{eqn: measurements with errors}).

 We call a set of  buses and lines as \textit{target buses} and \textit{target lines} respectively if the intruder attempts to change the congestion pattern of these lines by injecting errors on the target buses. Since an intruder may have limited resources to launch attacks, we assume at each time step the intruder can manipulate the states of at most $d$ bus voltages. The intruder, therefore, has at most $\sum_{i=0}^{d}\tbinom{M}{i}$ ways to select target buses.

The launched attacks can be detected by some recently developed methods, as presented in section \ref{sec:problem}. Here we use $p_d(s,a)$ to denote the probability that an action $a$ at state $s$ is detected by the network operator. We suppose it is a function of injected errors on the bus voltage magnitudes and angles:
%The detection probability $P_D(s,a)$ is defined  as a function of injected errors on the states of target lines:
\begin{equation}\label{eqn: detection probability}
p_d(s,a)=1-\exp\left(-C\times\sum_{i=1}^{M}\left(\frac{|e_{V_i}|}{\Delta V_i}+\frac{|e_{\theta_i}|}{\Delta \theta_i}\right)\right),
\end{equation}
where $C$ is a positive constant. Intuitively, a larger $C$ means a higher probability with which the launched attack can be detected.

Since the bus voltage magnitudes and angles are in discretized states, instead of computing the power flow of line $ij$ directly from one specific state $(\bar{\bm{V}},\bar{\bm{\theta}})$, we can obtain the lower and upper bound of its absolute value, denoted as $P_{ij}^{\min}(\bar{\bm{V}},\bar{\bm{\theta}})$ and $P_{ij}^{\max}(\bar{\bm{V}},\bar{\bm{\theta}})$ respectively.
Since the reward results from a change in the congestion pattern, we define the reward as a function proportional to the gap between the line's flow limit and the power bounds with injected errors:
\begin{equation}\label{eqn: attack reward}
R_{ij}=
\begin{cases}
K_{ij}\times \left( P_{ij}^{\min}(\bar{\bm{V'}}, \bar{\bm{\theta'}})-P_{ij}^{\text{M}} \right)/P_{ij}^{\text{M}}, \\
\text{if}\enspace P_{ij}^{\min}(\bar{\bm{V'}}, \bar{\bm{\theta'}})>P_{ij}^{\text{M}}>P_{ij}^{\max}(\bar{\bm{V}},\bar{\bm{\theta}}); \\
K_{ij}\times \left( P_{ij}^{\text{M}}-P_{ij}^{\max}(\bar{\bm{V'}}, \bar{\bm{\theta'}}) \right)/P_{ij}^{\text{M}}, \\
\text{if}\enspace P_{ij}^{\min}(\bar{\bm{V}}, \bar{\bm{\theta}})>P_{ij}^{\text{M}}>P_{ij}^{\max}(\bar{\bm{V'}},\bar{\bm{\theta'}}).		
\end{cases}
\end{equation}
where $K_{ij}$ is the given reward weight of line $ij$, $P_{ij}^{\text{M}}$ is the power flow limit of line $ij$, $(\bar{\bm{V'}},\bar{\bm{\theta'}})$ is the resulting estimate of system states by injecting errors $(\bm{e_V},\bm{e_\theta})$ to $(\bar{\bm{V}},\bar{\bm{\theta}})$.

%Following formula (\ref{eqn: real power}), the lower and upper bounds are computed as follows
%\begin{equation}
%P_{\min}^{ij}=
%\begin{cases}
%\frac{\text{lb}(V^i)\times \text{lb}(V^j)}{Z_{ij}}\sin(\text{lb}(\theta_i)-\text{ub}(\theta_j)) & \text{if}\  \theta^i>\theta^j\\
%0	&  \text{if}\ \theta^i=\theta^j\\
%\frac{\text{lb}(V^i)\times \text{lb}(V^j)}{Z_{ij}}|\sin(\text{ub}(\theta_i)-\text{lb}(\theta_j))| &  \text{if}\ \theta^i<\theta^j\\
%\end{cases}
%\end{equation}
%\begin{equation}
%P_{\max}^{ij}=
%\begin{cases}
%\frac{\text{ub}(V^i)\times \text{ub}(V^j)}{Z_{ij}}\sin(\text{ub}(\theta_i)-\text{lb}(\theta_j)) & \text{if}\  \theta^i>\theta^j\\
%0	&  \text{if}\ \theta^i=\theta^j\\
%\frac{\text{ub}(V^i)\times \text{ub}(V^j)}{Z_{ij}}|\sin(\text{lb}(\theta_i)-\text{ub}(\theta_j))| &  \text{if}\ \theta^i<\theta^j\\
%\end{cases}
%\end{equation}
%where lb(*) and ub(*) is the lower bound and upper bound of state variable in its corresponding value interval respectively.

%The detection probability $P_D(s,a)$ increases when the injected errors increase. 
%where $C$ is a positive tuning parameter and ; $\Phi(a)$ is the index set of target lines. Formula (\ref{eqn: detection probability}) follows that the detection probability increases with the increase of injected errors. 
The expected immediate reward from action $a$ at state $s$ is:
\begin{equation}\label{eqn: expected reward}
\begin{split}
R(s,a) & =\sum_{s'}P(s'|s,a)\times R(s'|s,a)\\
& =(1-p_d(s,a))\times \sum_{ij\in \Phi(s,a)}R_{ij}.
\end{split}
\end{equation}
where $\Phi(s,a)$ is the set of target lines. 

We assume the cost to intrude an accessible  measuring device is fixed and known to an intruder, denoted by $g_u$. Let $f(\Phi(s,a))$ denote the number of intruded measuring devices in attack $a$, the attack cost at state $s$ is:
\begin{equation}\label{Formula: attack cost}
G(s,a)=g_u\times f(\Phi(s,a)).
\end{equation}

\subsubsection{State Transition Probabilities}
%The  transition probabilities of PMU states are illustrated in  Fig.~\ref{state transition of PMU}. 
We assume all measuring devices that are currently open to attack  will stay open without attack. An action $a$ at state $s$ is detected  with probability $p_d(s,a)$, and the intruded  devices will be protected as a whole in the next time step.  Each protected device will change to open in the next time step with a fixed probability $p_T$. %$P_T$ here is used to reflect the protection response to attack, since 
Intuitively,  a smaller $p_T$ indicates that once protected, a device is more likely to stay inaccessible to intruders for a longer period of  time.  When $p_T=0$, it means the protected devices will no longer be accessible to intruders.

To model the dynamics of a power system, we assume each load in the system evolves independently as a Markov Chain \cite{MP94} and the system state is determined from economic dispatch. We assume each load has $n_L$ states and a load can transit from state $q_1$ to state $q_2$ with a fixed and known probability $p_{q_1,q_2}$. In practice, one can learn these transition probabilities from historical data. In this case, in a power network with $M$ buses and $N$ devices, if $M_L$ loads evolve as Markov Chains, the total number of system state is $n_L^{M_L}\times2^N$.

\subsection{MDP Solutions}\label{sec: MDP solution}
%Numerous algorithms have been developed to solve MDP, like  . 
A linear programming approach \cite{FR03,MP94} is applied to solve MDP. A stationary policy $\pi$ for an MDP is a mapping $\pi: S\mapsto A$, where $\pi(s)$ is the action taken in state $s$. We define $W_\pi(s)$ as the cumulative expected net reward by starting from state $s$ and following policy $\pi$, 
\begin{equation}
W_\pi(s)=E \left[\sum_{t=0}^{\infty}\gamma^t\left(R(s_t,\pi(s_t))-G(s_t,\pi(s_t))|s_0=s\right)\right],
\end{equation}
where $\gamma \in [0,1)$ is the discount factor for future reward.
The value of state $s$ is the maximal cumulative reward, 
\begin{equation}\label{Formula: maximal accumulated reward}
W^*(s):=\max_{\pi\in \Pi}\,W_\pi(s),
\end{equation} 
where $\Pi$ is the set of all available policies. The policy that achieves the maximum in (\ref{Formula: maximal accumulated reward}) is the optimal policy $\pi^*$.  It is shown in  \cite{FR03} that  $W^*(s)$ is the optimal solution to  the following optimization problem:
\begin{equation}\label{Formula: initial LP}
	\begin{split}
		\min_Q & \sum_{s\in S} Q(s)\\
		\text{s.t.}\ & Q(s)\geq R(s,a)-G(s,a)+\gamma \sum_{s'} P(s'|s,a)Q(s')\\ & \qquad \qquad \forall a\in A(s),\ \forall s,s' \in S.
	\end{split}
\end{equation}
Therefore, we can find $W^*(s)$ by solving   (\ref{Formula: initial LP}) 
and compute $\pi^*(s)$ defined as 
\begin{equation}\label{eqn:astar}
\argmax_{a\in A(s)}(R(s,a)-G(s,a)+\gamma \sum_{s'} P(s'|s,a)W^*(s')).
\end{equation}
 The optimal attack strategy is a mapping between system state $s$ and the corresponding optimal action $\pi^*(s)$. An intruder can solve the MDP to obtain the strategy offline and then inject attacks accordingly in real-time operations.
%We define $W(s)$ as the accumulated expected net reward of state $s$: 
%\begin{equation}
%W(s)=\sum_{t=0}^{\infty}\gamma^t(R(s_t,a)-G(s_t,a)|s_0=s),
%\end{equation}
%and $W^*(s)$ as the maximal accumulated reward function:
%\begin{equation}\label{Formula: maximal accumulated reward}
%W^*(s)=\max_{a\in A}\,W(s).
%\end{equation} 
%The action that achieves the maximum in (\ref{Formula: maximal accumulated reward}) is the optimal action for state $s$, denoted by $a^*(s)$. It is shown in \cite{FR03} that $W^*(s)$ is the optimal solution to (\ref{Formula: initial LP}), so we can find $W^*(s)$ by solving the linear program of (\ref{Formula: initial LP}) 
%and the optimal actions can be computed based on the function:
%\begin{equation}
%a^*(s)=\arg\max_a(R(s,a)-G(s,a)+\gamma \sum_{s'} P(s'|s,a)W^*(s')).
%\end{equation}

\section{Simulation}\label{sec:simu}
\begin{figure}[h]
	\centering
	\includegraphics[width=0.37\textwidth]{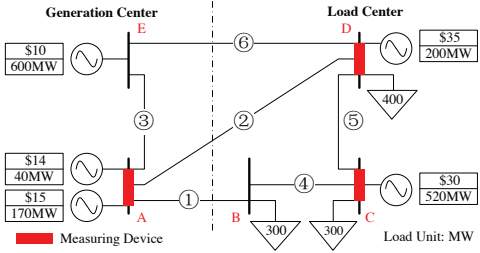}
	\caption{The PJM 5-bus system } \vspace{-3mm}
	\label{fig: PJM 5-bus system}
\end{figure}

%In this section we present one simulations in which the proposed model is applied and (\ref{Formula: initial LP}) is solved to get the optimal attack policy. The number of lines under attack $d$ is one and the power network is shown in Fig.~\ref{Simple example of medium power network}.
We test our proposed method on the PJM 5-bus system. The basic system configuration and the generation bids, generation MW limits and MW loads are shown in Fig.~\ref{fig: PJM 5-bus system}. More details can be found in \cite{LR10}.
\vspace{-2mm}
%There are two PMUs installed in the network, PMU 1 records the measurements of $V_3, I_{31}, I_{32}$ and $I_{35}$, and PMU 2 records the measurements of $V_5, I_{53}$ and $I_{54}$. 

\noindent\textit{1) States of loads and transition probability.}
Each load is assumed to have 2 states. The loads in Fig.~\ref{fig: PJM 5-bus system} are the base case loads. Another state for each load is half of its base case. The transition probability from one state to another is 0.5. 

\noindent\textit{2) Measuring device transition probability.}
$p_T$ is set 0.5. The attack detection probability is calculated from (\ref{eqn: detection probability}).

\noindent\textit{3) Costs.}
 The number of target buses at each time step is at most one. The reward weight of each line is 1. The cost $g_u$ is set as 0.05.

\noindent\textit{4) Discount factor.}
The discount factor $\gamma=0.95$.

We solve the economic dispatch in MATPOWER toolbox in MATLAB. The power flow limit of each line is set 300 MW. We relax the constraint in economic dispatch from $P_{ij}\leq P_{ij}^{\text{M}}$ to $P_{ij}\leq 1.2 P_{ij}^{\text{M}}$. We set $V^{\max}=1.1$ p.u. (per unit), $V^{\min}=1.0$ p.u., $n_V=5$; $\Delta \theta=5\degree, n_{\theta}=10, \theta_i^{\max}$ and $\theta_i^{\min}$ are determined from the actual values of bus $i$ in eight load states.  For each discretized state, we calculate the lower and upper bounds of each line power flow. One line can be an available target line if the upper bound of its power flow is below the power limit or the lower bound is over the limit.

\begin{figure}[h]
	\centering
	\includegraphics[width=0.37\textwidth]{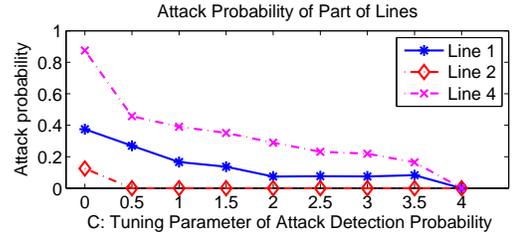}
	\caption{Attack probabilities of part of lines } \vspace{-3mm}
	\label{fig: Attack probabilities of lines}
\end{figure}

\begin{figure}[h]
	\centering
	\includegraphics[width=0.37\textwidth]{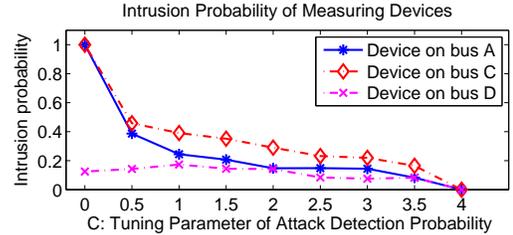}
	\caption{Intrusion probabilities of measuring devices } \vspace{-3mm}
	\label{fig: Intrusion probabilities of devices}
\end{figure}

We solve (\ref{Formula: initial LP}) and (\ref{eqn:astar}) to obtain the optimal actions for $2^3\times 2^3=64$ states and compute the static distribution probabilities of all states. The attack probability of one line is computed as the sum of the distribution probabilities of the states under which the optimal action is to change the congestion pattern of that line. The intrusion probability of one device is the sum of the distribution probabilities of the states under which the optimal action requires manipulating partial or all measurement of that device. The results of the likelihood of data attacks are shown in Figs.~\ref{fig: Attack probabilities of lines} and \ref{fig: Intrusion probabilities of devices}.

\vspace{-2mm}
\vspace{-3mm}

Generally, as $C$ increases, the attack probability of each line and each measuring device decreases. When $C=0$, it means the launched attack cannot be detected by network operators. In this case, the intruders always choose the action that brings about the maximal immediate net reward. When $C\geq 4$, the net expected reward for each action is negative, hence the optimal action for all states is no attack.  

\section{Conclusion and Discussions}\label{sec:con}
We for the first time analyze the likelihood of cyber data attacks to power systems. We model an intruder's attack strategy as a Markov Decision Process (MDP). We compute the optimal attack action at a given power system state from an intruder's perspective. We study the likelihood of cyber data attacks on a small system through simulation.
One ongoing work is to apply the method to likelihood analysis of cyber data attacks on larger power systems.
%Since the computational time of solving an MDP may be exponential in the system size, we are   implementing approximation algorithms to solve the MDP. 
%% In this paper, we analyze cyber data attack from an intruder's prospective and model the process of launching attacks on power system as a Markov decision process. Different from other scholar's work, in this model we take into consideration the influence of data attack on PMUs' state across time and model the dynamics of power system in the framework of Markov chains. The reward for intruders to launch attacks is from the price gap between the under-capacity and over-capacity states of transmission lines. The problem is solved by linear programming to find the optimal policy to launch attacks that yields the maximal rewards in the infinite horizon.
%
%In the simulation, we apply our model in one simple power network and get the optimal action for each state. Further we analyze the likelihood and frequency of cyber data attack on each line and study the influence of detection performance of attack identification algorithms on the attack likelihood. 
%
%Though our work is studied from an intruder's viewpoint, actually it can help system operators to defend the attack. For example, with the attack likelihood analysis on each line, to some extent it can help operators to identify the protection priority of PMUs in the system.  

\section*{Acknowledgement}
%We thank New York Power Authority for providing PMU data for the Central NY Power System. %This work is supported in part by NSF/DOE CURENT ERC and NYSERDA.
This research is supported in part by the ERC %Engineering Research Center
Program of NSF and DoE under the supplement to 
NSF Award   EEC-1041877 and the CURENT Industry Partnership
Program, and in part by NYSERDA Grants  \#36653 and \#28815. 
%The research is supported by NSF under CCF-0835706, ONR under N00014-11-1-0131, and DTRA under HDTRA1-11-1-0030.
\bibliographystyle{IEEEtranS}
%\bibliography{../bibfiles/IEEEabrv,../bibfiles/ref,../bibfiles/MengWangPub}
%\bibliography{./bibfiles/IEEEabrv,./bibfiles/ref,./bibfiles/MengWangPub}
% \input{appendix}
\end{document}